\catcode`\@=11
 
\newif\if@fewtab\@fewtabtrue
{\count255=\time\divide\count255 by 60
\xdef\hourmin{\number\count255}
\multiply\count255 by-60\advance\count255 by\time
\xdef\hourmin{\hourmin:\ifnum\count255<10 0\fi\the\count255}}
\def\ps@draft{\let\@mkboth\@gobbletwo
    \def\@oddhead{}
    \def\@oddfoot{\hbox to 7 cm{\tiny \versionno
       \hfil}\hskip -7cm\hfil\rm\thepage \hfil {\tiny\draftdate}}
    \def\@evenhead{}\let\@evenfoot\@oddfoot}
\def\draftdate{\number\month/\number\day/\number\year\ \ \ \hourmin }

\global\def\draftcontrol{0}
 
\def\draftcite#1{\ifnum\draftcontrol=1#1\else{}\fi}
\def\@lbibitem[#1]#2{\item{}\hskip -3\hbox to 2cm
{\hfil$\scriptstyle\draftcite{#2}$}\hskip
1cm[\@biblabel{#1}]\if@filesw
     {\def\protect##1{\string ##1\space}\immediate
      \write\@auxout{\string\bibcite{#2}{#1}}}\fi\ignorespaces}
 
\def\@bibitem#1{\item\hskip -3cm \hbox to 2cm
{\hfil {\footnotesize\draftcite{#1}}}\hskip 1cm
\if@filesw \immediate\write\@auxout
       {\string\bibcite{#1}{\the\value{\@listctr}}}\fi\ignorespaces}
 
\def\citen#1{\if@filesw \immediate\write \@auxout {\string\citation{#1}}\fi%
\@tempcntb\m@ne \let\@h@ld\relax \def\@citea{}%
\@for \@citeb:=#1\do {\@ifundefined {b@\@citeb}%
    {\@h@ld\@citea\@tempcntb\m@ne{\bf ?}%
    \@warning {Citation `\@citeb ' on page \thepage \space undefined}}%
    {\@tempcnta\@tempcntb \advance\@tempcnta\@ne
    \setbox\z@\hbox\bgroup\ifcat0\csname b@\@citeb \endcsname \relax
    \egroup \@tempcntb\number\csname b@\@citeb \endcsname \relax
    \else \egroup \@tempcntb\m@ne \fi \ifnum\@tempcnta=\@tempcntb
    \ifx\@h@ld\relax \edef \@h@ld{\@citea\csname b@\@citeb\endcsname}%
    \else \edef\@h@ld{\hbox{--}\penalty\@highpenalty
    \csname b@\@citeb\endcsname}\fi
    \else \@h@ld\@citea\csname b@\@citeb \endcsname \let\@h@ld\relax \fi}%
\def\@citea{,\penalty\@highpenalty\hskip.13em plus.13em minus.13em}}\@h@ld}
\def\@citex[#1]#2{\@cite{\citen{#2}}{#1}}%
\def\@cite#1#2{\leavevmode\unskip\ifnum\lastpenalty=\z@\penalty\@highpenalty\fi%
  \ [{\multiply\@highpenalty 3 #1%
  \if@tempswa,\penalty\@highpenalty\ #2\fi}]}   %
\makeatother 

\catcode`\@=12

\def\cald  {{\cal D}}

\def\calh  {{\cal H}}

\def\calw  {{\cal W}}

\def\dl            {\bf}

\def\zet           {{\dl Z}}

\long\def\query#1{\hskip 0pt{\vadjust{\everypar={}\small\vtop to 0pt{\hbox{}%
     \vskip -13pt\rlap{\hbox to 49.0pc{\hfil{\vtop{\hsize=8pc\tolerance=6000%
     \hfuzz=.5pc\rightskip=0pt plus 3em\noindent#1}}}}\vss}}}}%

\newcommand\nsection[1]   {\bigskip \noindent{\bf#1} \bigskip}
\def\be            {\begin{equation}}
\def\bearl         {\begin{array}{l}}
\def\bearll        {\begin{array}{ll}}

\def\cft           {conformal field theory}
\def\cfts          {conformal field theories}

\def\ee            {\end{equation}}

\def\eear          {\end{array}}

\newcommand\erf[1] {(\ref{#1})}

\def\futnote#1     {\footnote{~#1}\ }
\def\g             {{\bf g}}

\def\hy            {$\mbox{-\hspace{-.66 mm}-}$}

\def\irrep         {irreducible representation}
\def\kma           {Kac\hy Moody algebra}
\long\def\labl#1   {\label{#1}\ee \ifnum\draftcontrol=1
                   \mbox{ }\\[-12 mm]\query{#1}\\[5 mm] \fi}
\long\def\Labl#1#2 {\label{#1#2}\ee\ifnum\draftcontrol=1
                   \mbox{ }\\[-12 mm]\query{#1#2}\\[5 mm] \fi}

\def\wzwts         {WZW\hy theories}

\def\Ad         {{\rm Ad}}

\def\gb         {\bar\g}
\def\gp         {\g'}
\def\gbp        {\gb'}

\def\h          {{\bf h}}
\def\hb         {\bar{\h}}

\documentclass[12pt]{article} \usepackage{amssymb,amsfonts}
\begin{document}


\begin{flushright}  {~} \\[-1cm] {\sf hep-th/9908185} \\[1mm]
{\sf ETH-TH/99-21} \\[1 mm]
{\sf August 1999} \end{flushright}
 
\begin{center} \vskip 22mm
{\Large\bf New maverick coset theories}\\[22mm]
{\large Bill Pedrini} , \, 
{\large Christoph Schweigert}, 
\ and \ {\large Johannes Walcher}\\[5mm]
Institut f\"ur Theoretische Physik \\
ETH H\"onggerberg \\[.2em] CH -- 8093~~Z\"urich
\end{center}
\vskip 26mm
\begin{quote}{\bf Abstract}\\[1mm]
We present new examples of maverick coset conformal field theories. They 
are closely related to conformal embeddings and exceptional modular 
invariants.

\end{quote}
\vskip 20mm


\nsection{Introduction}

The coset construction \cite{baha,goko} provides a powerful tool
to construct explicit examples of \cfts. It has found numerous applications:
in superstring theory coset \cfts\ with extended supersymmetry have
been used to construct string compactifications \cite{kasu2,sche3,fuSc},
and recently the use of the coset construction has been proposed to relate
different \cft\ descriptions of the fractional Quantum Hall Effect
(\cite{cagt,frst}, and references therein, see also \cite{fpsw}).

The structure of coset theories turns out to be amazingly rich. It is
by now well understood that in the correct treatment of coset \cfts\
group theoretical selection rules lead to field identifications, and that
field identification fixed points require a special procedure called
{\em fixed point resolution} \cite{scya5,fusS4}. 

The phenomenon of field identification can be traced back to the following
fact: group theoretical selection rules for the embedding 
\be (\g')_{Ik} \hookrightarrow (\g_k) \ee
typically imply that certain branching functions vanish and that non-vanishing
branching functions coincide. (Here subscripts denote the level, $k$ is an
integer and $I$ is the Dynkin index of the embedding.)
In particular, there are different realizations
of states of conformal weight zero in the coset theory. All these states
should be identified to provide a theory with a unique vacuum. It turns
out that this can be done using simple current techniques \cite{scya5}.

It came as a surprise that group theoretical selection rules are {\em not}
the only reason for the existence of additional copies of the vacuum. 
First examples of this phenomenon were exhibited in \cite{dujo} and
\cite{dujo2}. Following these references, we will call {\em maverick} coset 
theory any coset \cft\ which has a non-vanishing branching function of
conformal weight $\dot\Delta=0$ that is {\em not} a consequence of group
theoretical selection rules. An ADE-classification for mavericks was proposed 
in \cite{dujo2}; however, one more maverick coset theory was found in 
\cite{fusS4}. The
known mavericks share the following features (see more below): they appear at
level $k=2$, and for level $1$ the embedding is a conformal embedding
\cite{scwa,babo}. In all cases the coset theory can be cured by use of
exceptional modular invariants.

The purpose of this note is twofold. We first exhibit a different class of 
maverick coset \cfts\ which have rather different features, but are still 
closely related to exceptional modular invariants. Secondly, we present
a new maverick coset at level $k=2$ and make some general remarks on this
type of maverick coset theories.

\nsection{New maverick coset theories at level $k=1$}

We start with the embedding 
\be  (A_1)_{10} \hookrightarrow (A_3)_1 \, , \ee
induced by the four-dimensional representation of $A_1$. The corresponding
coset \cft\ has the identification current $(J^2/J')$ of order two and
central charge $c=1/2$. Hence, if there is a \cft\ corresponding to this
coset theory, it can only be the critical Ising model. Applying the standard
procedure of field identification \cite{scya5,fusS4} to the theory in question
one obtains 11 primary fields; the computation
\footnote{The computations in this paper have partly been performed using 
the computer program {\sc kac} written by Bert Schellekens. {\sc kac}
is available at {\tt http://norma.nikhef.nl/\~{}t58/kac.html}.}
of the conformal weights $\dot\Delta$ for the branching functions 
$\Phi^\lambda_{\lambda'}$ (where $\lambda$ is an 
integrable highest weight of $(\g)_k=(A_3)_1$ and $\lambda'$ is an integrable
highest weight of the subalgebra $(\g')_{Ik}=(A_1)_{10}$) gives the
following values for $\dot\Delta\bmod \zet$:
\be\begin{array}{lllll}
\Phi^{000}_0 \cong \Phi^{010}_{10} & \dot\Delta=0 &\qquad &
     \Phi^{100}_{1} \cong \Phi^{001}_{9} & \dot\Delta=5/16\\
\Phi^{000}_2 \cong \Phi^{010}_{8} & \dot\Delta=5/6 &&
     \Phi^{100}_{3} \cong \Phi^{001}_{7} & \dot\Delta=1/16\\
\Phi^{000}_4 \cong \Phi^{010}_{6} & \dot\Delta=1/2 &&
     \Phi^{100}_{5} \cong \Phi^{001}_{5} & \dot\Delta=31/48 \\
\Phi^{000}_6 \cong \Phi^{010}_{4} & \dot\Delta=0 &&
     \Phi^{100}_{7} \cong \Phi^{001}_{3} & \dot\Delta=1/16\\
\Phi^{000}_8 \cong \Phi^{010}_{2} & \dot\Delta=1/3 &&
     \Phi^{100}_{9} \cong \Phi^{001}_{1} & \dot\Delta=5/16\\
\Phi^{000}_{10} \cong \Phi^{010}_{0} & \dot\Delta=1/2 && &
\end{array}\ee

The conformal weights expected for the Ising model ($0,
1/2, 1/16)$ appear twice, and moreover conformal weights appear that
do not show up in the Kac table for the Ising model. Using fermionization,
it is not hard to show that the following character identities hold:
\be\begin{array}{lll}
\chi_{000}(\tau) &=& 
\dot\chi_0(\tau) \,  \left(\chi'_0(\tau) + \chi'_6(\tau)\right) \, + \,
\dot\chi_{1/2}(\tau)  \, \left(\chi'_4(\tau) + \chi'_{10}(\tau)\right) \\
\chi_{010}(\tau) &=& 
\dot\chi_0(\tau)  \, \left(\chi'_{10}(\tau) + \chi'_4(\tau)\right) \, + \,
\dot\chi_{1/2}(\tau)  \, \left(\chi'_6(\tau) + \chi'_{0}(\tau)\right) \\
\chi_{100}(\tau) = \chi_{001}(\tau) &=& 
\dot\chi_{1/16}(\tau)  \, \left(\chi'_{3}(\tau) + \chi'_7(\tau)\right)
\end{array}\ee
Here $\dot\chi_{\dot\Delta}$ denote characters of the Ising model,
$\chi_\lambda$ 
characters of the ambient algebra $\g=A_3$ and $\chi'_{\lambda '}$ of the
subalgebra $\g'=A_1$. We observe that those branching functions that have
conformal weights that do not appear in the Kac table are indeed vanishing,
although they are perfectly allowed by the group theoretical selection
rules. Moreover, each Ising primary is realized by 
different branching functions that are {\em not} related by the action
of simple currents. In particular, there are more representatives of the
vacuum, $\Phi^{000}_6 \cong \Phi^{010}_{4}$, so that the coset is clearly
maverick.

However, there is indeed an {\em exceptional} modular invariant that is exactly
suited to make sense of the coset: at level $k=10$ the chiral algebra of $A_1$ 
allows for an exceptional extension which leads to an exceptional
modular invariant of extension type, the so-called $E_6$ exceptional modular 
invariant in the ADE-classification of modular invariants \cite{caiz2}. 

Another coset which exhibits similar features is provided by the embedding
\be (A_1)_{28} \hookrightarrow (B_3)_1 \, , \ee
which has an identification current $(1/J')$ of order two. Since it has
central charge $c=7/10$, it should be the tricritical Ising model. It
is straightforward to check that in this case the extension leading to the
$E_8$-modular invariant of $A_1$ at level $k=28$ has to be used.

What is the general pattern behind these mavericks? Suppose that the
embedding $(\gp)_{Ik}\hookrightarrow (\g)_k$ allows an intermediate
algebra $\h$, 
\be (\gp )_{Ik} \hookrightarrow (\h)_k \hookrightarrow (\g)_k \,, \labl{form}
such that the embedding $(\gp)_{Ik} \hookrightarrow (\h)_k$ is  
a conformal embedding at level $k=1$. We will show that in this case the
coset theory  
$\g/\gp$ is maverick at level $k=1$ if and only if the conformal embedding 
does not correspond to an extension by integer spin simple currents.

In the two examples, the relevant conformal embeddings are $(A_1)_{10}
\hookrightarrow (B_2)_1=(C_2)_1$ and $(A_1)_{28}
\hookrightarrow (G_2)_1$, respectively. Other examples, e.g.\ for $\g'=A_2$,
are easily 
obtained from the conformal embeddings
\be \begin{array}{lll}
(A_2)_5 &\hookrightarrow & (A_5)_1 \\
(A_2)_9 &\hookrightarrow & (E_6)_1 \\
(A_2)_{21} &\hookrightarrow & (E_7)_1 
\end{array}\ee
which correspond to the exceptional extensions of the chiral algebra of
$A_2$ that give rise to modular invariants of $A_2$ of extension type.

We now present an explicit argument which shows that all cosets of the 
form \erf{form} contain non-vanishing branching functions of conformal weight 
zero. To this end we consider the decomposition of the adjoint representation
of the horizontal subalgebra $\hb$ of $\h$ into $\gbp$ \irrep s.
This decomposition certainly contains the adjoint representation of $\gbp$, 
and at least one more \irrep. Let $\lambda'$ be the highest weight of one such 
\irrep. We claim that in the coset $\g/\gp$ the branching function for
$\Phi^\Omega_{\lambda'}$, where $\Omega$ is the vacuum of $\g$, is 
non-vanishing and has conformal weight zero. 

The conformal weight $\dot\Delta_{\lambda/\lambda'}$  of a coset primary field 
$\Phi^\lambda_{\lambda'}$ is computed according to the formula
\be \dot\Delta_{\lambda/\lambda'}= \Delta_\lambda-\Delta_{\lambda'} + n 
\,, \ee
where $n$ is the lowest degree in the affine module $\calh_\lambda$ on which
the $\gp$-module $\calh_{\lambda'}$ appears. In the present situation,
we compute $n$ by decomposing the vacuum $\calh_\Omega$ of the $\g$-theory
in terms of $\gp$ modules. The lowest degree contains a single state and
therefore only contributes to the vacuum $\calh_{\Omega'}$ of the $\gp$-theory.
On degree one, the adjoint of $\gb$ appears. In its decomposition in 
$\hb$-modules the adjoint of $\hb$ appears. Decomposing the adjoint
representation of $\hb$ further into $\gbp$ modules, we find $\lambda'$
among the $\gbp$ modules. This module therefore appears at degree one,
$n=1$, and moreover, we see that the branching function for 
$\Phi^\Omega_{\lambda'}$ is non-vanishing. Since $\lambda=\Omega$ is the
vacuum, we obviously have $\Delta_\lambda=0$.

It remains to compute $\Delta_{\lambda'}$. This is by definition the eigenvalue
of the zero mode of the Virasoro algebra obtained by the affine Sugawara
construction for $\gp$. However, since the embedding $\gp\hookrightarrow \h$
is conformal, we can equally well use the zero mode of the Virasoro algebra
for $\h$ to determine the conformal weight. In terms of $\h$ modules,
however, the relevant state appears at degree 1 of the vacuum module and
therefore has conformal weight $\Delta_{\lambda'}=1$. We conclude that the
coset conformal weight is $\dot\Delta_{\lambda/\lambda'}= 0-1+1=0$.

So far our argument applies to any conformal embedding. Now we have to
consider two different situations: either $\lambda'$ is a simple current of
the $\gp$-theory or not. The first case happens when the conformal embedding
$(\gp)_I\hookrightarrow(\h)_1$ is described by an integer spin simple current 
extension of $\gp$; the identification current $(\Omega/\lambda')$
will then do the appropriate field identification, according to the standard
prescription \cite{scya5,fusS4}.

The second case happens when the  embedding does not correspond to a simple
current extension. Then clearly the standard prescription does 
not identify the branching function with vanishing conformal
weight with the vacuum of the coset theory. The coset is therefore maverick.
However, the coset still yields a consistent conformal field theory,
we only have to use the exceptional extension of the chiral algebra of the
$\gp$-theory that is provided by the conformal embedding.

Coset \cfts\ also have a description as gauged \wzwts. Field identification
and fixed point resolution that are due to group theoretical selection 
rules can be traced back in this language to the fact that a non-simply
connected quotient of the real compact connected Lie group corresponding to
the embedded algebra has to be gauged \cite{hori}. 
In this language, the above situation can be described as the rather puzzling
observation that at level $k=1$ the subgroup corresponding to $\gp$ cannot
be consistently gauged. Rather, one has inevitably to gauge the larger
subgroup corresponding to $\h$ to obtain a consistent coset theory.
A deeper understanding of this fact in a Lagrangian formulation is not
known to us; we would like to emphasize that this effect exclusively
occurs at level $k=1$, but not at higher level.

Remarkably enough, the present class of theories shows
an intimate relation between maverick coset theories, exceptional extensions
of the chiral algebra (and therefore exceptional modular invariants) and 
conformal embeddings, although here these relations are
much less mysterious than in the coset theories of \cite{dujo,dujo2}.

\pagebreak

\nsection{Some remarks on maverick coset theories at $k=2$}

The mechanism we have presented for maverick coset theories at level
$k=1$ does not extend to the maverick cosets of 
\cite{dujo,dujo2} nor does it provide any insight in the classification
of mavericks of this type. In fact, a first counterexample to the conjectured 
ADE-classification \cite{dujo2} of mavericks was provided in
\cite{fusS4}. One more example is provided by the coset theory
\be (D_6)_2 \oplus (A_1)_2 \hookrightarrow (E_7)_2 \labl{bm}
which has $c=4/5$ and turns out to be the tetracritical Ising model.
It is therefore in particular a counterexample to the conjecture of
\cite{dujo2} that maverick cosets have extended $\calw$-symmetry.

The coset theory \erf{bm} shares the following four features of the mavericks
of \cite{dujo2,fusS4}. \\
1) The ambient algebra $\g$ is simply laced and at level $k=2$; at level $k=1$ 
   the embedding is a conformal embedding. \\
2) The pair $(\g^c,\gp^c)$ is a symmetric Lie algebra of compact type, where
$\g^c$ and $\gp^c$ stand for the compact real forms of $\g$ and $\gp$, respectively.\\
3) Identification currents occur that are trivial for the ambient algebra $\g$,
   and the branching functions of vanishing conformal weight $\dot\Delta=0$ 
   that are not consequences of group theoretical selection rules are fixed 
   points under these identifications. \\
4) Finally, and most remarkably, the following observation of J.\ Fuchs 
\cite{jfpriv} is valid also in this maverick: consider the fixed point 
$\Phi^{\lambda_0}_{\lambda'_0}$ that has conformal weight $\dot\Delta=0$ and
hence has to be identified with the vacuum. For the quantum dimensions 
of the $\g$-primary $\lambda_0$ and the $\g'$-primary $\lambda'_0$ one
has the relation
\be \cald_{\lambda_0} = \frac12 \cald_{\lambda_0'}. \ee
We remark that the coset \erf{bm} does not share the property
of the mavericks of \cite{dujo,dujo2} that $\lambda_0=\theta$ is the adjoint
representation and that $\lambda'_0$ appears in the decomposition of the
adjoint representation of $\gb$ in $\gbp$ \irrep s.

The fourth property of maverick cosets seems to be a crucial feature of maverick
coset theories. For example, consider the coset
\be (A_5)_2\oplus (A_1)_2 \hookrightarrow (E_6)_2 \ee
with central charge $c=25/28$. It shares the first three properties of 
mavericks. While it has an identification current of the form $(1/J'^3J')$ which does 
have fixed points, the ratio of the quantum dimensions of the $\g$ and $\gp$
primaries of the fixed points is 
in this case not exactly equal to two, but rather equals
\be \frac{\sqrt2}{\sqrt{2-\sqrt2} \, \cos(\pi/7)} \simeq 2.050858 \, . \ee
One can convince oneself that this coset is in fact {\em not} a maverick.

\medskip

The present examples of maverick coset theories show that a general 
understanding of field identification in coset theories still requires further 
clarification. 

\vskip2em\noindent{\small {\bf Acknowledgement} \\
We would like to thank J.\ Fr\"ohlich and J.\ Fuchs for discussions.}
 
 \vskip3em
\small
 \newcommand\wb{\,\linebreak[0]} \def\wB {$\,$\wb}
 \newcommand\Bi[1]    {\bibitem{#1}}
 \newcommand\J[5]   {{\sl #5}, {#1} {#2} ({#3}) {#4} }
 \newcommand\PhD[2]   {{\sl #2}, Ph.D.\ thesis (#1)}
 \newcommand\Prep[2]  {{\sl #2}, preprint {#1}}
 \newcommand\BOOK[4]  {{\em #1\/} ({#2}, {#3} {#4})}
 \def\jf    {J.\ Fuchs}
 \newcommand\inBO[7]  {in:\ {\em #1}, {#2}\ ({#3}, {#4} {#5}), p.\ {#6}}
 \newcommand\inBOx[7] {in:\ {\em #1}, {#2}\ ({#3}, {#4} {#5})}
 \newcommand\gxxI[2] {\inBO{GROUP21 Physical Applications and Mathematical
              Aspects of Geometry, Groups, and \A s{\rm, Vol.\,2}}
              {H.-D.\ Doebner, W.\ Scherer, and C.\ Schulte, eds.}
              \WS\Si{1997} {{#1}}{{#2}}}
 \def\anop  {Ann.\wb Phys.}
 \def\comp  {Com\-mun.\wb Math.\wb Phys.}
 \def\duke  {Duke\wB Math.\wb J.}
 \def\duki  {Duke\wB Math.\wb J.\ (Int.\wb Math.\wb Res.\wb Notes)}
 \def\foph  {Fortschr.\wb Phys.}
 \def\ijmp  {Int.\wb J.\wb Mod.\wb Phys.\ A}
 \def\imrn  {Int.\wb Math.\wb Res.\wb Notices}
 \def\jams  {J.\wb Amer.\wb Math.\wb Soc.}
 \def\jopa  {J.\wb Phys.\ A}
 \def\jstp  {J.\wb Stat.\wb Phys.}
 \def\jhep  {J.\wb High\wB Energy\wB Phys.}
 \def\mpla  {Mod.\wb Phys.\wb Lett.\ A}
 \def\nupb  {Nucl.\wb Phys.\ B}
 \def\phlb  {Phys.\wb Lett.\ B}
 \def\phrd  {Phys.\wb Rev.\ D}
 \def\phrl  {Phys.\wb Rev.\wb Lett.}
 \def\thmp  {Theor.\wb Math.\wb Phys.}
 \newcommand\geap[2] {\inBO{Physics and Geometry} {J.E.\ Andersen, H.\
            Pedersen, and A.\ Swann, eds.} \MD\NY{1997} {{#1}}{{#2}} }
 \def\AMS    {{American Mathematical Society}}
 \def\AP     {{Academic Press}}
 \def\CUP    {{Cambridge University Press}}
 \def\MD     {{Marcel Dekker}}
 \def\NH     {{North Holland Publishing Company}}
 \def\SV     {{Sprin\-ger Ver\-lag}}
 \def\WS     {{World Scientific}}
 \def\Ad     {{Amsterdam}}
 \def\Be     {{Berlin}}
 \def\Ca     {{Cambridge}}
 \def\NY     {{New York}}
 \def\PR     {{Providence}}
 \def\Si     {{Singapore}}

\end{document}